\documentclass[a4paper,11pt]{article}
\usepackage[dvipsnames]{xcolor}
\usepackage[utf8]{inputenc}
\usepackage[bbgreekl]{mathbbol}
\usepackage{geometry}
\usepackage{makecell}

\usepackage{xcolor}
\usepackage{amsmath, array, amssymb, amsfonts,amsthm}
\usepackage{upgreek}
\usepackage{xfrac}
\usepackage[inline]{enumitem}
\setlist{nolistsep}
\usepackage[french, english]{babel}
\usepackage[all]{xy}
\usepackage{txfonts}  % good1
\usepackage{sectsty}
\usepackage{booktabs}
\usepackage{caption}
\usepackage{dsfont}
\usepackage{mathtools}
\usepackage{slashed}
\usepackage[makeroom]{cancel}
\usepackage[hidelinks]{hyperref}
\usepackage{textcomp}
\usepackage{multicol}
\usepackage[title]{appendix}
\usepackage[square,numbers,sort,compress,semicolon,merge]{natbib}
\let\cite\citep % \cite is then \citep of natbib
\usepackage{hyperref}
\hypersetup{colorlinks=true, urlcolor=blue, citecolor=blue, linktoc=page}
\usepackage{tikz}
\usepackage{tikz-cd}
\usetikzlibrary{cd}
\tikzcdset{
arrow style=tikz,
diagrams={>={Straight Barb[scale=0.8]}}
}
\usetikzlibrary{matrix,arrows,decorations.pathmorphing}

\DeclareMathAlphabet{\mathpzc}{OT1}{pzc}{m}{it}

\usepackage{fancybox}
\usepackage{mathrsfs}
\usepackage{scrextend}
\usepackage{mathrsfs}
\usepackage{footmisc}

\usepackage{mathtools}

\makeatletter
\renewcommand*\env@matrix[1][\arraystretch]{%
  \edef\arraystretch{#1}%
  \hskip -\arraycolsep
  \let\@ifnextchar\new@ifnextchar
  \array{*\c@MaxMatrixCols c}}
\makeatother

\newcommand{\defeq}{\vcentcolon=}
\newcommand{\rdefeq}{=\vcentcolon}

\newcommand\M{\mathcal{M}}

\newcommand\RR{\mathbb{R}}

\newcommand\C{\mathcal{C}}

\renewcommand\H{\mathcal{H}}

\newcommand\E{\mathcal{E}}

\renewcommand\L{\mathcal{L}}
\renewcommand\S{\mathcal{S}}

\newcommand\SU{\mathcal{SU}}
\newcommand\U{\mathcal{U}}
\newcommand\SO{\mathcal{SO}}

\newcommand\K{\mathcal{K}}
\newcommand\J{\mathcal{J}}
\renewcommand\O{\mathcal{O}}

\newcommand\D{\mathcal{D}}

\newcommand\n{\text{\tiny{N}}}

\renewcommand\epsilon{\varepsilon}

\newcommand\rarrow{\rightarrow}

\newcommand\LieG{\mathfrak{g}}
\newcommand\LieH{\mathfrak{h}}

\newcommand\so{\mathfrak{so}}

\renewcommand\b{\bar }

\renewcommand\d{\partial}
\newcommand\s{\sigma}
\newcommand\bs{\boldsymbol}

\renewcommand\-{^{-1}}

\newcommand\ad{\text{ad}}

\makeatletter

\newcommand{\Rmnum}[1]{\expandafter\@slowromancap\romannumeral #1@}
\makeatother

\makeatletter
\newcommand{\leqnomode}{\tagsleft@true\let\veqno\@@leqno}
\newcommand{\reqnomode}{\tagsleft@false\let\veqno\@@eqno}
\makeatother

\DeclareMathOperator{\Diff}{Diff}

\DeclareMathOperator{\vol}{vol}

\theoremstyle{definition}

\makeatother

\begin{document}

\title{Dressing fields for supersymmetry: \\ The cases of the Rarita-Schwinger and gravitino fields}
\author{J. François${\,}^{a,\,b,\,c}$, L. Ravera${\,}^{d,\,e,\,f}$}
\date{}

\maketitle
\begin{center}
\vskip -0.8cm
\noindent
${}^a$ Department of Mathematics \& Statistics, Masaryk University -- MUNI. \\
Kotlářská 267/2, Veveří, Brno, Czech Republic. \\[2mm]
${}^b$ Department of Philosophy -- University of Graz. \\
Heinrichstraße 26/5, 8010 Graz, Austria. \\[2mm]
${}^c$ Department of Physics, Mons University -- UMONS.\\
Service \emph{Physics of the Universe, Fields \& Gravitation}. \\
20 Place du Parc, 7000 Mons, Belgium. \\[2mm]
${}^d$ DISAT, Politecnico di Torino -- PoliTo. \\
Corso Duca degli Abruzzi 24, 10129 Torino, Italy. \\[2mm]
${}^e$ Istituto Nazionale di Fisica Nucleare, Section of Torino -- INFN. \\
Via P. Giuria 1, 10125 Torino, Italy. \\[2mm]
${}^f$ \emph{Grupo de Investigación en Física Teórica} -- GIFT. \\
Universidad Cat\'{o}lica De La Sant\'{i}sima Concepci\'{o}n, Concepción, Chile. %\\[2mm]
\end{center}

\vspace{-3mm}

\begin{abstract}
In this paper we argue that the gauge-fixing conditions typically used to extract the (off-shell) degrees of freedom of the Rarita-Schwinger spinor-vector and gravitino, respectively in rigid supersymmetric field theory and supergravity, are actually instances of the dressing field method of symmetry reduction. Since the latter has a natural relation interpretation, solving the ``gauge-fixing condition" -- or, better, ``dressing functional constraints" -- actually realises the Rarita-Schwinger spinor-vector and the gravitino fields as (non-local) relational variables. 
To the best of our knowledge, this is the first application of the dressing field method to supersymmetric theories.

\end{abstract}


%%%%%%%%%%% MAIN TEXT %%%%%%%%%%%%%%%%%
\vspace{-3mm}

\tableofcontents

\bigskip

%\clearpage

\section{Introduction}
\label{Introduction}

Gauge-fixing is an ubiquitous procedure in field theory, both  classically  and in the quantization of gauge theories. 
In supersymmetric field theories, it is typically considered as the key tool allowing to write down the off-shell number of degrees of freedom (d.o.f.) of the fields\footnote{Obviously, one should care to distinguish off-shell d.o.f. from on-shell d.o.f.:  the latter being, arguably, the physical ones. 
It frequently happens that for some authors the spin of the field is meant as a proxy to the physical -- on-shell -- d.o.f. (e.g. $A_\mu$ has spin $1$, ergo, 2 d.o.f.), whereas for others the spin is meant to identify the index structure of the field, and therefore refer to the off-shell degrees of freedom.
In the latter case, one further distinguishes the number of free components implied by the index structure from the implicit gauge-invariant d.o.f. counted by quotienting out the relevant gauge symmetries (e.g. off-shell $A_\mu$ has a priori 4 free components, but only 3 when quotienting out the $\U(1)$ gauge symmetry).} involved in the supermultiplet, in particular those of the Rarita-Schwinger (RS) spinor-vector \cite{Rarita:1941mf} in the flat case,
and of the gravitino in supergravity -- where, as we shall see, the most common gauge choice is a gamma-tracelessness functional condition on the latter field.
Another possible gauge choice in this context relies on the transverse-longitudinal decomposition of the field, where the longitudinal part -- namely its divergence -- is set to zero. 
The former choice is usually preferred as an off-shell, i.e. kinematic, gauge-fixing in supersymmetric field theories to get the physical d.o.f., 
as there is no natural way to obtain it as a consequence of the dynamics, i.e. of the field equations derived from a Lagrangian. 
%especially since it is harder to think of a Lagrangian theory in which it may emerge as a consistency condition following from the field equations. 
Furthermore, for the standard Lagrangian of the RS field, the divergencelessness constraint 
%This is the case, on the other hand, for the functional constraint removing the divergence of the field, which in fact 
follows as a consistency condition from the field equations and the gamma-tracelessness constraint   taken together -- see, e.g., \cite{Valenzuela:2022gbk,Valenzuela:2023aoa} for details on this point. 

As is well-known, imposing gauge-fixing constraints amounts to restricting one's attention to a subspace of the initial field space of a theory, and therefore restricting the original symmetry group to a subgroup preserving that subspace. 
The soundness of the motivations behind these restrictions may vary:  from mere matters of convenience, to  more ``foundational" issues, such as locating within the initial field space the d.o.f. one feels are the relevant ones for the purpose at hand. 
In any case, such motivations -- often only tacit -- must be made plain. 

We here highlight an alternative to gauge-fixing, which turns out to be relevant for the latter issue as it appears at the very foundation of supersymmetric field theory and supergravity: 
The so-called dressing field method (DFM) is a systematic geometric approach to the construction of  gauge-invariant variables,  first introduced in \cite{GaugeInvCompFields}. 
%and further developed  e.g.  in\cite{Francois2018,Francois2021,Francois2023-a} -- see refs. therein, also Chapter 5 of \cite{Berghofer-et-al2023}, and \cite{Zajac2023}. 
It is best understood within the formalism of the bundle differential geometry of field space \cite{Francois2021,Francois2023-a} (see also \cite{Zajac2023}), but has also a simple field-theoretic framing -- see e.g. Chapter 5 of \cite{Berghofer-et-al2023}. The DFM has a natural \emph{relational} interpretation: 
Gauge-invariance is achieved by extracting the \emph{physical} d.o.f. representing \emph{relations} among field variables \cite{Francois2023-a,JTF-Ravera2024c}.

It turns out that many instances of ``gauge-fixing constraint", when solved explicitly for the ``transformation parameter", actually produce a \emph{dressing field}. 
The corresponding ``gauge-fixed" field is actually a \emph{gauge-invariant dressed field} -- analogous to a Dirac dressing \cite{Dirac55, Dirac58} -- which means that it does not belong to the initial field space, and that the intended ``gauge-fixing" is therefore not a gauge-fixing in any mathematically meaningful sense. For~more on this key point, see \cite{Francois-Berghofer2024}. 

As we show here, the above mentioned gamma-tracelessness and divergencelessness ``gauge-fixing constraints" of supersymmetry and supergravity fall in that category. 
One thus produces, via the DFM, the RS spinor-vector and the gravitino as supersymmetry-invariant dressed fields, naturally interpreted as carrying the relational d.o.f. of the theory. 
In the latter, all fields are to be similarly dressed via the extracted dressing field. Those invariants dressed fields satisfy a set of dressed field equations.

\medskip
The paper is structured as follows: 
In section \ref{Symmetry reduction via dressing} we shortly review both the basics of gauge field theory and the DFM of symmetry reduction, highlighting in what the latter differs from gauge-fixing. Moreover, we  present for the first time the case of \emph{perturbative dressing} (or \emph{linear dressing}). 
In section  
\ref{Supersymmetry and dressing: The case of the Rarita-Schwinger field} we discuss the case of the RS spinor-vector field, commonly referred to as a spin-$\sfrac{3}{2}$ field, as a dressed object -- actually, we will see that, after dressing, it will contain both a spin-$\sfrac{3}{2}$ and a spin-$\sfrac{1}{2}$ fields, in terms of irreducible spin representations. 
Subsequently, in section \ref{Supergravity: The gravitino as a dressed field} we move on to supergravity, showing that the gravitino field is a (perturbatively) dressed field. 
In both the RS and the gravitino cases we analyse two different functional constraints to extract the dressing field: the gamma-tracelessness condition and the divergencelessness condition. We discuss both the kinematic and the dynamics of the (dressed) theories. 
We sum-up and conclude in section \ref{Conclusion}, while in appendix \ref{Dressing superfield in superspace} we give our proposal of a possible dressing superfield in the ``geometric" (\emph{rheonomic} \cite{Castellani:1991eu}) approach to supersymmetric field theories in superspace.

\section{Symmetry reduction via dressing}
\label{Symmetry reduction via dressing}

In this section, after introducing the typical field content and the associated geometric structure of classical gauge field theory, we briefly review the dressing field method of symmetry reduction, stressing its difference from the gauge-fixing procedure.

\subsection{Basics of gauge theory}
\label{Field content and geometry of gauge theories}

The basic fields of a gauge theory based on a Lie group $H$ with Lie algebra $\LieH$, over the region $U\subset M$ of a $n$-dimensional manifold $M$, are the gauge potential -- or connection -- 1-form $A=A^a={A^a}_\mu\, dx^{\,\mu} \in \Omega^1(U, \LieH)$ and the matter fields $\phi \in \Omega^\bullet (U,V)$, with $V$ a representation space via $\rho:H \rarrow GL(V)$, and $\rho_*:\LieH \rarrow \mathfrak{gl}(V)$. Their minimal coupling is given by the covariant derivative $D\phi := d\phi + \rho_*(A)\, \phi \in \Omega^{\bullet+1}(U,V)$. The field strength of $A$, i.e. the curvature, is $F=dA+\sfrac{1}{2}[A,A] \in \Omega^2(U,\LieH)$.

These are subject to the action of the (infinite-dimensional) gauge group $\H$ of the theory, that is the set of $H$-valued functions  $\upgamma: U \rightarrow H$, $x \mapsto \upgamma(x)$, with
point-wise group multiplication $(\upgamma \upgamma')(x)=\upgamma(x) \upgamma'(x)$, defined  by the property that  any given $\upeta \in \H$, seen as a ``field" on $U\subset M$,  is acted upon by any other $\upgamma \in \H$ as  $\upeta \mapsto \upgamma^{-1} \upeta\upgamma=:\upeta^\upgamma$.
The gauge group is then by  definition
 \begin{align}
 \label{Gauge-group}
\H := \left\{ \upgamma, \upeta :U \rightarrow H\ |\  \upeta^\upgamma\defeq \upgamma^{-1} \upeta\upgamma\, \right\}.
\end{align} 
Its Lie algebra is thus 
\begin{align}
\label{LieGaugeGrp}
\text{Lie}\H := \Big\{ \lambda, \lambda' :U \rightarrow \LieH\ |\  \delta_{\lambda}\lambda' \defeq [\lambda', \lambda]\, \Big\},
\end{align}
where 
$\delta_{\lambda}\lambda'$ denotes the action of $\lambda$ on~$\lambda'$. 
The~action of $\H$ (or Lie$\H$) defines the gauge transformations
\begin{equation}
\begin{aligned}
\label{GTgauge-fields}
&A\ \mapsto\ A^\upgamma:=\upgamma^{-1} A \upgamma + \upgamma^{-1} d \upgamma , \quad \phi \ \mapsto \ \phi^\upgamma:=\rho(\upgamma)^{-1}\phi,\\[1mm]
&\text{infinitesimally, } \quad \delta_\lambda A = D\lambda=d\lambda +\ad(A) \lambda, \quad \delta_\lambda \phi=-\rho_*(\lambda)\, \phi,
\end{aligned} 
\end{equation}
which imply
$F\ \mapsto\ F^\upgamma=\upgamma^{-1} F \upgamma$
and
$D\phi \mapsto (D\phi)^\upgamma :=\ d\phi^\upgamma + \rho_*(A^\upgamma)\phi^\upgamma =\rho(\upgamma)^{-1}D\phi$.  
Defining the field space of the theory as $\Phi=\{A, \phi\}$, by \eqref{Gauge-group}-\eqref{GTgauge-fields} the action of the gauge group $\H$ on  $\Phi$ is a right action: $(A^\upeta)^\upgamma=A^{\upeta\upgamma}$ and $(\phi^\upeta)^\upgamma=\phi^{\upeta\upgamma}$.
E.g. for $\phi$, omitting $\rho$, one has indeed: \mbox{$(\phi^\upeta)^\upgamma\!=(\upeta\- \phi)^\upgamma = (\upeta^\upgamma)\- \phi^\upgamma = (\upgamma\- \upeta \upgamma)\- \upgamma\- \phi = \upgamma\-\upeta\- \phi =(\upeta\upgamma)\-\phi \rdefeq~\!\!\phi^{\upeta\upgamma}$}.
\mbox{We~denote} the $\H$-gauge orbit of $\{A, \phi\}$ as $\O^\H_{\{A, \phi\}}\subset \Phi$.
%, or $\O_{\{A, \phi\}}$. 
The  action of $\H$ foliates $\Phi$ into  gauge orbits $\O^\H$ which, \mbox{under} \mbox{adequate} restrictions on either $\Phi$ (excluding field configurations with stability subgroups) or $\H$ (considering \ \mbox{elements} s.t.$\,\upgamma ={e_H}_{\,|\d U}$), are isomorphic~to~$\H$. 
See e.g. \cite{Singer1978, Singer1981, Ashtekar-Lewandowski1994, Baez1994,  Fuchs-et-al1994, Fuchs1995}.
Under these conditions, the field space $\Phi$ is a principal fiber bundle with structure group $\H$ over the moduli  space of orbits $\M\defeq\Phi/\H$. We have $\Phi \xrightarrow{\pi} \M$, with projection map $\pi(\{A, \phi\})=\{[A], [\phi]\}$. 

The dynamics of a gauge field theory is given by a Lagrangian form $L=L(A, \phi) = \L (A,\phi) \vol_n \in \Omega^n(U, \RR)$, where $\vol_n$ is the volume form on $U$, which is typically required to be quasi-$\H$-invariant, meaning $\H$-invariant up to boundary terms: $L(A^\upgamma, \phi^\upgamma)=L(A, \phi) + db(A, \phi; \upgamma)$, for $\upgamma\in \H$, so that the field equations $\bs E(A, \phi)=0$ remain $\H$-covariant. 

\paragraph{Gauge-fixing}
One often finds it convenient to make computations more manageable by restricting to those variables $\{A, \phi\}$ satisfying particular functional properties. 
When these restrictions are imposed by exploiting the gauge freedom \eqref{GTgauge-fields} of the field variables, we  call them ``\emph{gauge-fixing}" conditions. 
The functional restrictions are chosen so as to define a ``slice" in $\Phi$, cutting across gauge orbits once, selecting a single representative in each.

A gauge-fixing is but a \emph{choice of local section} of the field space bundle $\Phi$, i.e. $\s: \U\subset\M \rarrow \Phi$.
Concretely, it is  specified by a gauge condition taking the form of an algebraic and/or differential equation on the field variables %$\C(A, \phi)=0$. 
 using the gauge freedom \eqref{GTgauge-fields} explicitly: $\C(A^\upgamma, \phi^\upgamma)=0$. 
The~gauge-fixing slice is the submanifold 
%\begin{align}
%\label{GF}
$\S\defeq \{(A, \phi) \in \Phi_{|\U}\, |\, \C(A^\upgamma, \phi^\upgamma)=0 \} \subset \Phi$,
%\end{align}
and is  the image of the local section $\s$. 
There is no such global section, no ``good" gauge-fixing, unless  $\Phi$ is trivial, i.e. $\Phi=\M \times \H$. %\footnote{
Famously, the Gribov-Singer obstruction (or Gribov ambiguity) is the statement that no global gauge-fixing exists for pure $\H=\SU(n)$-gauge theories over compact regions of spacetime \cite{Singer1978, Singer1981, Fuchs1995}.

\subsection{The dressing field method}
\label{Dressing field method in a nutshell}

We here provide the basics of the DFM. 
Consider a $\H$-gauge theory  with Lagrangian $L(A, \phi) \in \Omega^n(U, \RR)$. 
\vspace{-1mm}

\paragraph{Kinematics}
Suppose there is a subgroup $K \subseteq H$ to which corresponds the gauge subgroup $\K \subseteq \H$. 
Suppose also there is a group $G$ s.t. either  $H\supseteq G \supseteq K$, or  $G \supseteq H$. 
A $\K$-\emph{dressing field}  is a map $u: M \rarrow G$, i.e. $G$-valued field, defined by its $\K$-gauge transformation: 
\begin{align}
\label{GT-dressing}
u^\kappa:=\upkappa\- u, \quad \text{ for } \upkappa \in \K.
\end{align} 
We denote the space of such $G$-valued $\K$-dressing fields by  $\D r[G, \K]$. 
 One  calls $\K$ (or $K$) the \emph{equivariance group} of $u$, while $G$ is its \emph{target group}.
Given the existence of a $\K$-dressing field, we have that
for $\{A, \phi \}$ as above, one may define the  \emph{dressed fields}
\begin{align}
\label{dressed-fields}
A^u\defeq  u\- A u + u\- du  \quad \text{ and } \quad \phi^u\defeq u\- \phi.
\end{align}  
These are $\K$-invariant, as is clear  from \eqref{GTgauge-fields} --  just replacing $\upgamma \rarrow \upkappa \in \K$ -- and \eqref{GT-dressing}.
 When $u$ is a $\H$-dressing field, $\K=\H$, the dressed fields \eqref{dressed-fields} are  $\H$-invariant.
 The corresponding dressed curvature is $F^u=u\- Fu=dA^u+\sfrac{1}{2}[A^u, A^u]$, and satisfies the Bianchi identity $D^{A^u}F^u=0$, where the dressed covariant derivative is $D^u=d + \rho_*(A^u)$, so that $D^u\phi^u =\rho(u)\-D\phi=  d\phi^u + \rho_*(A^u)\phi^u$.
For the dressings $\{A^u, \phi^u\}$ to make sense when $G \supset H$,  one must assume that representations of $H$ extend to representations of~$G$.

Observe that a \emph{pure gauge potential} $A_0$, a.k.a. flat connection, is necessarily given by a $\H$-dressing field: $A_0=udu\-$. By  $u^\upgamma =\upgamma\- u$, $A_0$ indeed $\H$-transforms as a gauge potential \eqref{GTgauge-fields}. 
A flat gauge potential is not
expressible as $\upgamma d \upgamma\-$ with $\upgamma \in \H$:  given the  transformation properties \eqref{Gauge-group} of members of the gauge group,  $\upgamma d \upgamma\-$ does not $\H$-transform as a gauge potential. 
\vspace{-1mm}

\paragraph{Dynamics}
Consider the Lagrangian  form $L=L(A, \phi)$ of a $\H$-gauge theory, satisfying  quasi-$\H$-invariance.
Suppose that there is a $\K$-dressing field $u$ with target group $G\subseteq H$. 
Exploiting the quasi-invariance of $L$, we may define  the \emph{dressed Lagrangian} as the Lagrangian expressed in terms of the dressed fields \eqref{dressed-fields}:
\begin{equation}
\label{dressed-Lagrangian}
L(A^u, \phi^u)=L(A, \phi) + db(A, \phi; u).
\end{equation}
If $L$ is strictly $\H$-invariant, i.e. such that  $b=0$, 
it can be rewritten as $L(A, \phi)= L(A^u, \phi^u)$. 
%The latter is a theory whose gauge symmetry has been reduced, as the $\K$ gauge subgroup has been neutralised. 
In both cases, the field equations $\bs E(A^u, \phi^u)=0$ for the dressed field have the same functional expression as  $\bs E(A, \phi)=0$.
\vspace{-1mm}

\paragraph{Residual gauge symmetry}
If $K$ is a normal subgroup of $H$,
 $K \triangleleft H$, then $H/K\rdefeq J$ is a Lie group. 
Correspondingly, $\K \triangleleft \H$ and $\J =\H/\K$ is a subgroup of $\H$.
The dressed fields \eqref{dressed-fields} may then exhibit well-defined residual $\J$-gauge transformations. 
In particular, if the $\K$-dressing field transforms as $u^\upeta =\upeta\- u\, \upeta$ for $\upeta\in \J$,  
the dressed fields  are $\J$-gauge variables, satisfying \eqref{GTgauge-fields}  under the replacement $\upgamma \rarrow \upeta$. 
Therefore, $L(A^u, \phi^u)$ is a $\J$-theory. 
If $L$ is $\H$-invariant, so that it can be rewritten as $L(A, \phi)= L(A^u, \phi^u)$, meaning that it is not  a $\H$-theory but a $\J$-theory. 
\vspace{-1mm}

\paragraph{Field-dependent dressing fields}
A central idea within the DFM, is that a dressing field should be extracted from the field content of the theory, rather than introduced in an \emph{ad hoc} way. 
These \emph{field-dependent dressing fields} are functionals on $\Phi$:
\begin{equation}
\label{Field-dep-dressing}
\begin{aligned}
u\ \ :\ \  \Phi \ &\rarrow\  \D r[G, \K], \\
    \{A, \phi\} \  &\mapsto\  u=u[A, \phi] .
\end{aligned} 
\end{equation}
The original field variables $\{A, \phi\}$ encode physical d.o.f. in a redundant way, mixing them with non-physical pure gauge modes.
The dressed fields $\{A^{u[A, \phi]}, \phi^{u[A, \phi]}\}$ can be understood as a reshuffling of the d.o.f. of the original fields that eliminates (part or all of) the pure gauge modes. 
If $u$ is a $\H$-dressing field, the $\H$-invariant dressed fields faithfully represent the physical  d.o.f. embedded in the initial set of variables $\{A, \phi\}$.  

It should be stressed that, for field-dependent dressing fields \eqref{Field-dep-dressing}, the DFM has a natural \emph{relational} interpretation:
The fields $\{A^{u[A, \phi]}, \phi^{u[A, \phi]}\}$ represent the gauge-invariant physical \emph{relations} among  d.o.f. embedded in $\{A, \phi\}$, either among d.o.f. of $A$ and $\phi$ themselves -- considering ``self-dressings", $A^{u[A]}$ or $\phi^{u[\phi]}$ --  or between d.o.f. of $A$ and $\phi$ -- considering e.g. $A^{u[\phi]}$ or $\phi^{u[A]}$.
See \cite{JTF-Ravera2024c} which makes the case for relationality as the conceptual core of general-relativistic gauge field theory, and 
\cite{Francois2023-a}
for the extension of the DFM to general-relativistic theories, with diffeomorphisms $\Diff(M)$ as symmetry, and the associated relational interpretation.

\paragraph{Dressing \emph{vs} gauge-fixing}
Is is clear that $u \notin \K$, as seen from comparing the definitions  \eqref{Gauge-group} of the gauge group and  that \eqref{GT-dressing} of a dressing field.
Therefore, in spite of the formal analogy with \eqref{GTgauge-fields}, the dressed fields   \eqref{dressed-fields} are not gauge transformations.
In particular, $A^u$ is no more a gauge potential, being $\K$-invariant. Also, in case $G\supset H$, the dressed potential is $\LieG$-valued, $A^u \in \Omega(U, \LieG)$, not $\LieH$-valued like $A$, and $F^u \in \Omega^2(U, \LieG)$. 
This happens e.g. in the case of the gauge treatment of gravity -- via Cartan geometry, see e.g. \cite{JTF-Ravera2024review} for a recent review.
The dressed fields $\{A^u, \phi^u\}$ are not a point in the gauge $\K$-orbit $\O^\K_{\{A, \phi\}} \subset \O^\H_{\{A, \phi\}}$ of $\{A, \phi\}$. Therefore, $\{A^u, \phi^u\}$ must not be confused with a gauge-fixing of $\{A, \phi\}$, i.e. with a point on a gauge-fixing slice $\S$. 
Contrary to the action of $\H$ or any of its subgroups, and thus contrary to gauge-fixing $\Phi \rarrow \S \subset \Phi$, the dressing operation is not a mapping from field space $\Phi$ to itself, but a mapping from field space  to another mathematical space: the space of dressed fields, denoted $\Phi^u$.

% \begin{minipage}{0.5\textwidth}
% \vspace{0.2cm}
% \begin{equation*}
% %\label{GF-map}
% \begin{aligned}
% \text{Action of } \H\  : \ \quad\Phi\  &\rarrow\ \Phi, \\   
% \text{Gauge-fixing}\ : \ \quad \Phi\  &\rarrow \ \S \subset \Phi, 
% \end{aligned}
% \vspace{0.2cm}
% \end{equation*}
% \end{minipage}
% \hspace{-2cm}
% \begin{minipage}{0.5\textwidth}
% \vspace{0.5cm}
% \begin{equation*}
% %\label{Dressing-map}
% \begin{aligned}
%  \text{Dressing}\ : \ \quad \Phi\  &\rarrow\  \Phi^u, \\
%  \ \{A, \phi\} &\mapsto \{A^u, \phi^u\}.
% \end{aligned}
% \end{equation*}
% \vspace{0.2cm}
% \end{minipage}

Yet, there is a relation between gauge-fixing and dressing which is as follows.
Consider a complete symmetry reduction via a $\H$-dressing field; then
$\Phi^u$ can be understood as a \emph{coordinatisation} of the moduli space $\M$ -- or a local region $\U \subset \M$  over which the field-dependent dressing field $u: \Phi_{|\U} \rarrow \D r[G, \H]$ is defined. 
In such a case, there is  a one-to-one mapping $(\Phi_{|\U})^u \leftrightarrow \U \subset \M$, $\{A^u, \phi^u\} \leftrightarrow \{[A], [\phi]\}$, i.e. a coordinate chart.
A gauge-fixing section over the same region, $\s: \U   \rarrow \S \subset \Phi_{|\U}$, provides a one-to-one mapping $\U \leftrightarrow \S$. 
There is thus an isomorphism of spaces $(\Phi_{|\U})^u \simeq \S\subset \Phi_{|\U}$. 
Still,  $(\Phi_{|\U})^u \neq \S$ as mathematical spaces. 
Remark that while performing the dressing procedure allows to work with the physical d.o.f. in a coordinatisation of the moduli space, the latter is not accessible in any way to direct computations through gauge-fixing. 
Furthermore, as stressed above, the DFM has a relational interpretation that gauge-fixing cannot have. We thus observe that, due to their formal similarity, a manifestly relational reformulation $L(A^u, \phi^u)$ of a theory  may be unfortunately conflated with a gauge-fixed version. For a more exhaustive treatment of the differences between dressing and gauge-fixing we refer the reader to \cite{Francois-Berghofer2024}.

\paragraph{Perturbative dressing}
It may be that one is interested in invariance at first order, i.e. under the  infinitesimal gauge transformations
\eqref{GTgauge-fields}. 
In this case, one may look for an infinitesimal Lie$\K$-dressing field 
\begin{align}
\label{pert-dressing-field}
\upsilon: M \rarrow \LieG, 
 \quad \text{s.t. } \quad \delta_\lambda \upsilon \approx -\lambda \ \ \text{for}\ \  \lambda \in \text{Lie}\K,
\end{align}
%with defining Lie$\K$-transformation 
where in the defining transformation law, one is to neglect higher order terms, polynomials in $\lambda$ and $\upsilon$.
Then one may define the perturbatively dressed fields $\upphi^\upsilon\defeq \upphi + \updelta_\upsilon \upphi$, where $\upphi=\{A, \phi\}$ and $\updelta_\upsilon \upphi$ mimics the functional expression of the infinitesimal gauge  transformation $\delta_\lambda \upphi$ \eqref{GTgauge-fields} with the substitution $\lambda \rarrow \upsilon$.\footnote{Notice that $\updelta_\upsilon$ is in no sense a differential, and does not stand on its own as a notation.}
Explicitly,
\begin{equation}
\label{pert-dressed-fields}
\begin{aligned}
A^\upsilon\defeq&\ A+ \updelta_\upsilon A   \quad
&&\phi^\upsilon \defeq \phi+ \updelta_\upsilon \phi
\\
=& \, A+ D\upsilon,   &&\phantom{\phi...}= \phi -\rho_*(\upsilon)\,\phi.
\end{aligned}
\end{equation}
We may show that the above are $\K$-invariant at first order, writing
\begin{align}
\delta_\lambda \upphi^\upsilon 
= \delta_\lambda \phi + \updelta_{\delta_\lambda \upsilon} \upphi 
= \delta_\lambda \phi + \updelta_{-\lambda} \phi
= \delta_\lambda \phi - \delta_\lambda \phi \equiv 0,
\end{align}
where one is to neglect higher order terms, bilinear in $\lambda$ and $\upsilon$, that we might see as arising in the above computation from the term written, in the suggestive but unsound notation,  ``$\updelta_\upsilon \delta_\lambda \phi$".
For the gauge potential and matter field we have, respectively, using directly \eqref{pert-dressed-fields} 
%\begin{equation}
\begin{align*}
    & \delta_\lambda A^\upsilon 
    = \delta_\lambda  A + \delta_\lambda D\upsilon 
    = D\lambda +D (\delta_\lambda \upsilon) + [\delta_\lambda A, \upsilon] 
    =D\lambda + D(-\lambda) 
    + \cancelto{\,\text{\tiny{neglect}}}{[D\Lambda, \upsilon]}\!\!
    %= D\lambda - D\lambda 
    \equiv 0 , \\
    & \delta_\lambda \phi^\upsilon = 
    \delta_\lambda  \phi - \delta_\lambda \big( \,\rho_*(\upsilon)\, \phi \, \big)
    = -\rho_*(\lambda)\, \phi - \rho_*(\delta_\lambda \upsilon)\, \phi - \rho_*(\upsilon)\,\delta_\lambda \phi
    =-\rho_*(\lambda)\, \phi - \rho_*(-\lambda)\, \phi - \cancelto{\,\text{\tiny{neglect}}}{ \rho_*(\upsilon)\rho_*( -\lambda)\, \phi}\!\!\!
    %= -\rho_*(\lambda)\, \phi + \rho_*(\lambda)\, \phi 
    \equiv 0.
\end{align*}
%\end{equation}
For a quasi-invariant Lagrangian, s.t. 
$\delta_\lambda L(A, \phi)= d\beta(A, \phi; \lambda)$, one may write the perturbatively dressed version
\begin{align}
\label{pert-dressed-Lagrangian}
  L(A^\upsilon, \phi^\upsilon)\defeq L(A, \phi) + d\beta(A, \phi; \upsilon). 
\end{align}
The field equations $\bs E(A^\upsilon, \phi^\upsilon)=0$ are thus  $\K$-invariant at first order (in both $\lambda$ and $\upsilon$).
Naturally, \eqref{pert-dressed-fields}-\eqref{pert-dressed-Lagrangian} may be obtained as linearisations  of \eqref{dressed-fields}-\eqref{dressed-Lagrangian}.

\medskip
The above DFM framework applies to graded Lie groups (supergroups), hence to supersymmetric gauge field theories and supergravity. 
%We will consider in the next sections the case of the so-called Rarita-Schwinger field and gravitino 1-form field, showing how the notion of dressing actually appears at the very basis of supersymmetric field theories. 
It indeed appears at the very foundation of the latter topic, as we show in the next sections where we consider  the case of the so-called Rarita-Schwinger field and gravitino 1-form field.

\section{Supersymmetry and dressing: The case of the Rarita-Schwinger field}
\label{Supersymmetry and dressing: The case of the Rarita-Schwinger field}

We will now show that two of the most common gauge-fixing choices used as foundational tools to achieve the desired number of off-shell d.o.f. in supersymmetric field theory are actually dressings. 
We will start from the case of the so-called Rarita-Schwinger (RS) field in supersymmetric field theories, considered to be a spinor-vector ${\psi^\alpha}_{\!\mu}$, component of a spinor-valued 1-form field $\psi = \psi_\mu \, dx^{\,\mu} \in \Omega^1(U,\sf S)$, with $U \subset M$ a 4-dimensional manifold and $\sf S$ a (Dirac) spinor representation for the Lorentz group $S\!O(1,3)$.\footnote{Given a basis of $\sf S$, $\{e_\alpha\}$, we have $\psi_\mu = {\psi^\alpha}_{\!\mu} e_\alpha$. For notational convenience we will frequently omit the spinor index $\alpha$.
We assume the Majorana reality condition $\bar \psi = \psi^\dagger \gamma^0 = \psi^t C$ -- with $C$ the charge conjugation matrix s.t. $C^t=-C$ -- so that $\psi$ and $\bar \psi$ are not independent fields.} 
For the sake of simplicity, we will focus on the $\mathcal{N}=1$ case. However, our discussion applies to higher-dimensional, $\mathcal{N}$-extended supersymmetric theories as well.

Before proceeding with this analysis, let us clarify, following the lines of \cite{Valenzuela:2022gbk,Valenzuela:2023aoa}, some aspects concerning RS fields and Lagrangian(s).

\paragraph{Spinor-tensor Rarita-Schwinger field in supersymmetric theories}

The spinor-tensors intruduced by Rarita and Schwinger in the original work \cite{Rarita:1941mf} are of the kind ${\psi^\alpha}_{\,\mu_1\ldots \mu_l}={\psi^\alpha}_{\,(\mu_1\ldots \mu_l)}$ and are commonly said to describe spin-$(l + \sfrac{1}{2})$ spinor-tensor fields on $U \subset \RR^{4}$. They fulfill a (massive) Dirac equation $(\slashed{\d}+m){\psi^\alpha}_{\,\mu_1\ldots \mu_l}=0$, where $\slashed{\d}:=\gamma^{\,\mu}\d_{\mu}$, with $\gamma^{\,\mu}$ the Dirac gamma-matrices in $n=4$ spacetime dimensions, together with the Lorentz-irreducibility condition $\gamma^{\,\mu_1}{\psi^\alpha}_{\,\mu_1\ldots \mu_l}=0$ -- a.k.a. gamma-traceless condition. If the RS spinor is a solution of those equations with $m \neq 0$, then it also necessarily satisfies ${\psi^{\mu_1}}_{\,\mu_1 \ldots \mu_l}=0$, $\d^{\mu_1}\psi_{\,\mu_1 \ldots \mu_l}=0$.
The case $l=1$ correspond to the simplest RS spinor-vector field ${\psi^\alpha}_{\,\mu}$, which is typically said to carry spin $\sfrac{3}{2}$, for which the condition ${\psi^{\mu_1}}_{\,\mu_1 \ldots \mu_l}=0$ does not hold; this is the field we will be interested in. 

Both the Dirac and the gamma-traceless conditions are on-shell constraints (field equations) of the family of Lagrangians introduced by Rarita and Schwinger. However, in the context of supersymmetric field theories and, in particular, of supergravity, the Lagrangian referred to as RS term is the (massless) theory
\begin{align}
\label{RSLagr}
    L_{\text{RS}}(\psi) = \bar \psi \wedge \gamma_5 \gamma \wedge d \psi \quad \rarrow \quad
    \L_{\text{RS}}(\psi) = \epsilon^{\,\mu \nu \rho \sigma} \bar \psi_\mu \gamma_5 \gamma_\nu \d_\rho \psi_\sigma \,,
\end{align}
where $\gamma:=\gamma_\mu \, d x^{\,\mu}$ is the gamma-matrix 1-form, and whose field equations read
\begin{align}
\label{RSfieldeqs}
    \gamma_5 \gamma \wedge d \psi = 0 \quad \rarrow \quad
    \epsilon^{\,\mu \nu \rho \sigma} \gamma_5 \gamma_\nu \d_\rho \psi_\sigma = 0 .
\end{align}
Observe that, in fact, such a theory does not coincide with the Lagrangian and field equations originally proposed by Rarita and Schwinger. In particular, the massless limit of the above equations does not correspond to \eqref{RSfieldeqs} and the variation of \eqref{RSLagr} does not imply the gamma-traceless condition. 
The Lagrangian \eqref{RSLagr} is quasi-invariant under the gauge transformation
\begin{equation}
\begin{aligned}
\label{susygaugetrRS}
    & \psi \mapsto \psi^\upepsilon=\psi + d \upepsilon \quad \rarrow \quad \psi_\mu \mapsto \psi^\upepsilon_\mu=\psi_\mu + \d_\mu \upepsilon , \\
    & \text{infinitesimally, } \quad \delta_\epsilon \psi = d \epsilon \quad \rarrow \quad \delta_\epsilon \psi_\mu = \partial_\mu \epsilon ,
\end{aligned}
\end{equation}
where $\upepsilon=\upepsilon(x)$ is a spin-$\sfrac{1}{2}$ Majorana parameter, more precisely an element of the Abelian (additive) gauge group 
\begin{align}
\label{stranslation-gauge-grp}
\E\defeq\Big\{\upepsilon, \upepsilon'\!:\!U \rarrow T^{0|4}\, |\, \upepsilon^{\upepsilon'}=\upepsilon\, \Big\},
\end{align}
where $T^{0|4}\subset T^{4|4}$ is the \emph{supertranslation} subgroup  of the ($\mathcal{N}=1$) super-Poincaré group $sIS\!O(1,3) \defeq S\!O(1,3) \ltimes T^{4|4}$ \cite{Gursey1987,DeAzc-Izq}, and the boundary term is simply  $db(\psi; \upepsilon)=d(\bar \upepsilon \wedge \gamma_5 \gamma \wedge d \psi )$.\footnote{We are using the notation $\epsilon=\delta \upepsilon$.}

Let~us observe that the field ${\psi^\alpha}_{\!\mu}$, commonly said to contain a spin-$\sfrac{3}{2}$ and a spin-$\sfrac{1}{2}$ part, actually involves, in terms of \emph{irreducible} spin representations, a spin-$\sfrac{3}{2}$ component and two spin-$\sfrac{1}{2}$ parts: 
$(1\oplus 0)\otimes \sfrac{1}{2}=\sfrac{3}{2}\oplus \sfrac{1}{2}\oplus \sfrac{1}{2}$ (see e.g. \cite{VanNieuwenhuizen:1981ae}). 
The spin-$\sfrac{1}{2}$ fields appearing into this (non-local) irreducible decomposition correspond to the gamma-trace and the divergence of ${\psi^\alpha}_{\!\mu}$. 
In fact, the decomposition -- most commonly mentioned in the supersymmetry literature -- in terms of a spin-$\sfrac{3}{2}$ and a spin-$\sfrac{1}{2}$ field only, i.e.
\begin{align}\label{gammatracedec}
    {\psi^\alpha}_{\!\mu} (\uprho,\chi) := {\uprho^\alpha}_{\!\mu} + \gamma_\mu \, \chi^\alpha ,
\end{align}
is \emph{reducible}. In this reducible spin decomposition, 
$\chi^\alpha := 1/n \, \gamma^{\,\mu} \, {\psi^\alpha}_{\!\mu}$ is a spin-$\sfrac{1}{2}$ field -- we remind that $n$ denotes the number of spacetime dimensions. 

The field ${\uprho^\alpha}_{\!\mu}$, which is s.t. $\gamma^{\,\mu}\uprho_\mu =0$, contains both a ``longitudinal" (divergence-free) and a ``transverse'' modes, 
${\uprho^{\alpha|\text{L}}}_{\!\mu}$ and 
${\uprho^{\alpha|\text{T}}}_{\!\mu}$, corresponding, respectively, to 8 and 4 off-shell d.o.f., that is to a 
spin-$\sfrac{3}{2}$ and a spin-$\sfrac{1}{2}$ contribution. 
We will come back to these points later, when discussing  dressing in this context. 
Here, let us just recall that, in the supersymmetry literature, it is commonly said that the spin-$\sfrac{1}{2}$ part of $\psi$ can be eliminated by ``gauge-fixing", typically
\begin{align}
\label{gammatr}
    \gamma^{\,\mu} \psi_{\mu} = 0 ,
\end{align}
and that one is therefore left with a spin-$\sfrac{3}{2}$ field (12 off-shell d.o.f.). 
Actually, those 12 off-shell d.o.f. correspond to the ones carried by the residual spin-$\sfrac{3}{2}$ and spin-$\sfrac{1}{2}$ contributions after ``gauge-fixing". 
Besides, as we shall see in a while, the condition \eqref{gammatr} actually involves a dressing procedure and the (commonly called) RS field $\psi$ is therefore a dressed (super)field. 
Before showing this explicitly, let us recall that, taking together \eqref{RSfieldeqs} and \eqref{gammatr}, one also obtains the transversality condition
\begin{align}
\label{transvcond}
    \d^{\,\mu} \psi_\mu = 0 ,
\end{align}
which therefore follows only on-shell in the theory at hand. 
Hence, $\psi$ satisfies $\Box \psi = 0$, where $\Box:=\d^{\,\mu} \d_\mu$, and thus $\psi$ describes (on-shell) a spin-$\sfrac{3}{2}$ massless particle propagating in a Minkowski background.\footnote{Remark that the gauge freedom \eqref{susygaugetrRS} cannot be used again to fix also \eqref{transvcond}, as it has been used to fix \eqref{gammatr}. 
The latter equation is obtained at the kinematic level (off-shell), it is a ``gauge choice", while \eqref{transvcond} here is obtained by exploiting also the field equations \eqref{RSfieldeqs}, and is therefore an on-shell result.}
Indeed, in the \emph{flat} case, using $[\d_\mu,\gamma_\mu]=0$ and the properties of the gamma-matrices in four spacetime dimensions, it can be easily shown that the field equation \eqref{RSfieldeqs} implies $\slashed\d (\gamma^{\,\mu} \psi_\mu)-\d^{\,\mu}\psi_\mu=0$, which, using the ``gauge choice" \eqref{gammatr}, yields \eqref{transvcond}. On the other hand, the field equations \eqref{RSfieldeqs} together with the ``gauge choice" \eqref{transvcond}, yield the weaker constraint $\slashed\d (\gamma^{\,\mu}\psi_\mu)=0$.  

\subsection{Rarita-Schwinger gamma-trace dressing}
\label{Rarita-Schwinger gamma-trace dressing}

The starting point of our analysis is the RS field ${\psi^\alpha}_{\!\mu}$, which carries, in principle, 16 d.o.f. and, as we have recalled above, can be decomposed according with $(1\oplus 0)\otimes \sfrac{1}{2}=\sfrac{3}{2}\oplus \sfrac{1}{2}\oplus \sfrac{1}{2}$. Nevertheless, to immediately see which is the dressing field in this context, in this section we will consider the reducible gamma-trace decomposition \eqref{gammatracedec}. 
Under~\eqref{susygaugetrRS}-\eqref{gammatracedec} we have
\begin{equation}
\begin{aligned}
\label{chirhogaugetr}
    \chi & \,\mapsto\, \chi^\upepsilon= \chi + \tfrac{1}{n} \slashed\d \upepsilon , \\
    \uprho_\mu & \,\mapsto \,\uprho^\upepsilon_\mu = \uprho_\mu - \tfrac{1}{n} \gamma_\mu \slashed \d \upepsilon + \d_\mu \upepsilon .
\end{aligned}
\end{equation}
Let us consider the gamma-tracelessness constraint \eqref{gammatr} as a functional condition on the variable $\psi_\mu^u\defeq \psi_\mu+ \d_\mu u$ and solve it explicitly for  the parameter $u$:
\begin{equation}
\begin{aligned}
\label{gamma-tr-dressing}
    & \gamma^{\,\mu} \psi^u_\mu = \gamma^{\,\mu} (\psi_\mu + \d_\mu u) = 0 , \\
    & \Rightarrow \quad u[\psi] = - \slashed{\d}\- (\gamma^{\,\mu} \psi_\mu) = - n \slashed{\d}\- \chi.
\end{aligned}
\end{equation}
To assess if $u$ is an element of the gauge group, and thus the gamma-trace constraint \eqref{gammatr} a genuine gauge-fixing, or if it is a dressing field, we have to ascertain its gauge transformation. 
For \eqref{gammatr} to be a gauge-fixing, $u$ must be an element of $\E$: i.e. it must be gauge-invariant, $u[\psi]^\upepsilon:=u[\psi^\upepsilon]=u[\psi]$, because the gauge group $\E$ of supertranslation is Abelian \eqref{stranslation-gauge-grp}.
As a functional of $\psi$, under \eqref{chirhogaugetr}, $u$ gauge transforms as
\begin{align}
u[\psi]^\upepsilon \defeq u[\psi^\upepsilon] 
=  - n \slashed{\d}\- \chi^\upepsilon 
=- n \slashed{\d}\- \left(\, \chi + \tfrac{1}{n} \slashed\d \upepsilon \right) 
=- n \slashed{\d}\- \chi - \upepsilon
=u[\psi] - \upepsilon,
\end{align}
which is, in fact, the Abelian (additive) version of a dressing field transformation \eqref{GT-dressing}. 
Therefore, explicitly solving the gamma-trace constraint \eqref{gammatr} does not result in a gauge-fixing, but in a dressing.
One thus has the corresponding gauge-invariant dressed field \eqref{dressed-fields}, 
\begin{align}
\label{dressed-RS}
\psi_\mu^u\defeq&\, \psi_\mu+ \d_\mu u [\psi]
= \psi_\mu   - n \d_\mu  \slashed{\d}\- \chi ,
\end{align}
which, by construction, satisfies
$\gamma^{\,\mu}\psi^u_\mu \equiv 0$.
Applying the gamma-trace decomposition \eqref{gammatracedec} to $\psi^u_\mu$, we find
\begin{equation}
\begin{aligned}
\label{chirhogammatrdressed}
    \chi^u & = \chi + \tfrac{1}{n} \slashed\d u[\psi] = \chi + \tfrac{1}{n} (-n) \slashed{\d} \slashed{\d}\- \chi \equiv 0 , \\
    \uprho^u_\mu & = \uprho_\mu - \tfrac{1}{n} \gamma_\mu \slashed \d u[\psi] + \d_\mu u[\psi] = \psi_\mu^u.
\end{aligned}
\end{equation}
Observe that this dressing is \emph{non-local}. 
The dressing field is given of the gamma-trace component $\chi$, carrying 4 d.o.f.; the resulting \emph{dressed field} $\psi^u = \uprho^u$ %-- the latter being, indeed, s.t. $\gamma^{\,\mu} \uprho^u_\mu=\gamma^{\,\mu} \uprho_\mu=0$ (see the second line of \eqref{chirhogammatrdressed}), by definition -- 
carries 12 physical off-shell d.o.f. ($16-4=12$, as $\chi^u=0$ by definition) and it still contains both a spin-$\sfrac{3}{2}$ (8 d.o.f.) and a spin-$\sfrac{1}{2}$ (4 d.o.f.) component.

Notice that the 12 off-shell d.o.f. of the dressed RS field \eqref{dressed-RS} are obtained in a \emph{gauge-invariant} way, without any restriction on the gauge group. 
Whereas, if one is claiming to restrict to this 12 d.o.f. by imposing \eqref{gammatr}, one is in effect restricting to the field space, to the subspace $\psi_\mu \rarrow \rho_\mu$ by  \eqref{gammatracedec}, by restricting the gauge group to those elements s.t. $\slashed{\d}\upepsilon=0$ by the first line of \eqref{chirhogaugetr}.

Finally, observe that the above gamma-trace dressing, obtained via the functional constraint $(\gamma^{\,\mu})^\alpha_{\phantom{\alpha}\beta}\,{\psi^{\,\beta}}_{\!\mu}=0$, on a RS spinor-vector field ${\psi^\alpha}_{\!\mu}$, is  analogous to the  ``\emph{axial gauge}" for a vector gauge potential $A_{\mu}$ in QED: $n^{\,\mu}A_{\mu}=0$, 
with~$n^{\,\mu}$ a constant 4-vector -- see e.g.  \cite{Leibbrandt-Richardson1992}. 
We indeed claim that as one solves explicitly (as above) the axial constraint, one ends-up building  a (non-local) dressing field $u[A, n]$, which  allows to write down the $\U(1)$-invariant (self-)dressed field $A^u_\mu\defeq A_\mu+ \d_\mu u[A, n]$.

\subsection{Transverse dressing}
\label{Transverse dressing}

Using the template of the previous section, we now consider another functional constraint \eqref{transvcond}. 
Usually understood as a gauge-fixing, we show that when solved explicitly, it again yields a dressing field. 
To better appreciate this, we perform the following (\emph{reducible}, non-local) decomposition for the RS spinor-vector field:
\begin{align}
\label{TLdivergdec}
\psi_\mu = \psi_\mu^{\text{T}} + \psi_\mu^{\text{L}} \defeq \psi_\mu^{\text{T}} + \d_\mu [ \Box\- (\d^{\,\nu} \psi_\nu) ]  = \psi_\mu^{\text{T}} + \partial_\mu \kappa ,
\end{align}
where $\psi^\text{T}_\mu$ is the transverse component, and $\psi^\text{L}_\mu:=\d_\mu \kappa$ is the longitudinal (``pure gauge", i.e. $d$-exact) component. %-- corresponding to the divergence of $\psi_\mu$. 
The ``pre-potential" of the latter, $\kappa$, is a spin-$\sfrac{1}{2}$ field  carrying 4 d.o.f. off-shell.
Under \eqref{susygaugetrRS} we have
\begin{equation}
\begin{aligned}
\label{psiTkappagaugetr}
    \psi_\mu^{\text{T}} & \,\mapsto\, \big(\psi_\mu^{\text{T}}\big)^\upepsilon= \psi_\mu^{\text{T}} , \\
    \psi_\mu^{\text{L}} & \,\mapsto \,\big(\psi_\mu^{\text{L}}\big)^\upepsilon= \psi_\mu^{\text{L}} + \d_\mu \upepsilon , \quad \text{that is} \quad \kappa \,\mapsto \,\kappa^\upepsilon = \kappa + \upepsilon .
\end{aligned}
\end{equation}
Observe that $\psi_\mu^{\text{T}}$ is invariant under the gauge transformation \eqref{susygaugetrRS}.
Now, we consider \eqref{transvcond} as a functional condition on the variable $\psi^u_\mu:=\psi_\mu + \d_\mu u$ and solve it explicitly for $u$, namely
\begin{equation}
\begin{aligned}
\label{div-less-dressing}
    & \d^{\,\mu} \psi^u_\mu =  \d^{\,\mu} (\psi_\mu + \d_\mu u) = 0 , \\
    & \Rightarrow \quad u[\psi] = - \Box\- (\d^{\, \mu} \psi_\mu) = - \Box\- (\d^{\, \mu} \psi^\text{L}_\mu) = - \kappa .
\end{aligned}
\end{equation}
Now, as a functional of $\psi$, under \eqref{psiTkappagaugetr}, $u$ gauge transforms as
\begin{align}
u[\psi]^\upepsilon \defeq u[\psi^\upepsilon] 
=  - \kappa^\upepsilon 
=- \kappa - \upepsilon
=u[\psi] - \upepsilon,
\end{align}
which is the Abelian (additive) version of a dressing field transformation \eqref{GT-dressing}. Hence, explicitly solving the divergencelessness condition \eqref{transvcond} does not result in a gauge-fixing, but in a dressing. The corresponding  dressed field reads
\begin{align}
\label{div-dressed-field}
    \psi^u_\mu := \psi_\mu + \d_\mu u = %\big(\psi^{\text{T}}\big)^u_\mu := 
    \big( \psi^\text{T}_\mu + \d_\mu \kappa \big) -\d_\mu \kappa =
    \psi^{\text{T}}_\mu ,
\end{align}
and it is both gauge-invariant and  divergence-free by construction.
%, given just by the transverse component, which is gauge-invariant under \eqref{susygaugetrRS} -- by construction, it satisfies $\d^{\,\mu}\psi^u_\mu=0$. 
Applying the decomposition \eqref{TLdivergdec} to $\psi^u_\mu$, we find, indeed,
\begin{equation}
\begin{aligned}
\kappa^u & = \kappa + u \equiv 0, \\
\big(\psi^u_\mu\big)^\text{T} & = \psi^{\text{T}}_\mu=\psi_\mu^u.
\end{aligned}
\end{equation}
The dressing field is \emph{non-local}, and given in terms of $\kappa$, which carries 4 d.o.f.; the resulting \emph{dressed field} 
$\psi^u =\psi_\mu^\text{T}$ %-- s.t. $\d^{\,\mu}\big(\psi^{\text{T}}_\mu\big)^u=0$, by definition -- 
carries 12 physical off-shell d.o.f. and still contains both a spin-$\sfrac{3}{2}$ (8 d.o.f.) and a spin-$\sfrac{1}{2}$ (4 d.o.f.) component. Indeed, observe that the dressed field is still gamma-tracefull, $\gamma^{\,\mu} \psi^u_\mu \neq 0$. The 12 physical off-shell d.o.f. are obtained, once again, in a \emph{gauge-invariant} way, without any restriction on the gauge group.

Finally, let us remark that the functional constraint \eqref{transvcond} is entirely analogous to  the  ``\emph{Lorenz gauge}" for a vector gauge potential $A_{\mu}$ in QED: $\d^{\,\mu} A_{\mu}=0$. See \cite{Francois-Berghofer2024} for the proof that solving explicitly (as above) the Lorenz constraint, one ends up building  a (non-local) dressing field $u[A]$, which  allows to write down the $\U(1)$-invariant (self-)dressed field $A^u_\mu\defeq A_\mu+ \d_\mu u[A]$. 

\subsection{Dynamics of the dressed theory}
\label{Dynamics of the dressed theory}

According to the DFM  framework \eqref{dressed-Lagrangian}, with either dressing fields,
the Lagrangian $4$-form of the dressed theory is thus
\begin{equation}
\begin{aligned}
\label{dressed-RSLagr}
    L_{\text{RS}}(\psi^u) &= \bar \psi^u \wedge \gamma_5 \gamma \wedge d \psi^u 
     \\
    &=L_{\text{RS}}(\psi) + db(\psi; u) \\
    &=\bar \psi \wedge \gamma_5 \gamma \wedge d \psi + d(\bar u \wedge \gamma_5 \gamma \wedge d \psi ).
\end{aligned}
\end{equation}
In components, the dressed Lagrangian density is $\L_{\text{RS}}(\psi^u) = \epsilon^{\,\mu \nu \rho \sigma} \bar \psi_\mu^u \gamma_5 \gamma_\nu \d_\rho \psi^u_\sigma$.
The dressed field equations are
\begin{align}
\label{dressed-RSfieldeqs}
    \gamma_5 \gamma \wedge d \psi^u = 0 \quad \rarrow \quad
    \epsilon^{\,\mu \nu \rho \sigma} \gamma_5 \gamma_\nu \d_\rho \psi_\sigma^u = 0.
\end{align}
The Lagrangian \eqref{dressed-RSLagr} is now $\E$-invariant, because $\psi^u$ is a supersymmetry singlet ($\E$-invariant). 
Remark that the dressed field $\psi^u$ is a \emph{relational variable} \cite{JTF-Ravera2024c}: it represents the physical, invariant relations among the (off-shell) d.o.f. of $\psi$. 
The dressed field equations \eqref{dressed-RSfieldeqs} are \emph{deterministic}, meaning that once the initial conditions are specified,
they uniquely determine the evolution of the \emph{relational} d.o.f., represented by $\psi^u$.

Observe also that $\E$-invariance is obtained at the cost of the locality of the theory, hinting at the fact that the supersymmetry $\E$ is what is called a \emph{substantial} gauge symmetry. On the contrary, an \emph{artificial} symmetry is one that is eliminated without losing locality. See \cite{Francois2018} for a discussion of this crucial point.

Finally, we may notice that, if the dressing field $u[\psi]$ is given by the gamma-trace constraint \eqref{gamma-tr-dressing}, the dressed field equation \eqref{dressed-RSfieldeqs} automatically implies $\d^\mu \psi^u_\mu \equiv 0$. 
Wehereas, if $u[\psi]$ is given by the divergenceless constraint \eqref{div-less-dressing}, the dressed field equation \eqref{dressed-RSfieldeqs} yields the weaker condition $\slashed{\d}(\gamma^\mu \psi^u_\mu)\equiv 0$.

\section{Supergravity: The gravitino as a dressed field}%%%%%%%%%%%%%%%%%%%%%%%%%%%%%
\label{Supergravity: The gravitino as a dressed field}%%%%%%%%%%%%%%%%%%%%%%%%%%%%%

The minimal coupling with gravity,
described by the (Lorentz) spin connection ${\omega^a}_b \in \Omega^1\big(U, \so(1,3)\big)$ and the soldering (vielbein) 1-form $e^a \in \Omega^1\big(U, \RR^4\big)$,
is obtained via the substitution
\begin{equation}
\begin{aligned}
    \d_\mu & \,\mapsto\, D_\mu \defeq\d_\mu + \rho_*(\omega_\mu) , \\
    \gamma_\mu & \,\mapsto\, \gamma_a {e^a}_\mu ,
\end{aligned}
\end{equation}
with $D_\mu$ %:=D_\mu(\omega)$ 
is the $\SO(1,3)$-covariant derivative, and $\{\gamma_a\}$  the flat space gamma-matrices --  $\gamma_\mu$ in the previous~section. 
For a  spinor field $\uplambda$, we have the representation 
$\rho_*=\sfrac{1}{4} \, \gamma_{ac} \eta^{bc} \defeq \sfrac{1}{8}\,[\gamma_a, \gamma_c]\eta^{bc}$ so that
$D_\mu\uplambda := d \uplambda + \sfrac{1}{4} \, {\omega^{ac}}_\mu \gamma_{ac} \uplambda$. Thus, the $\SO(1,3)$-invariant RS Lagrangian in a gravitational background is then
\begin{align}
   L_{\text{RS}}(\psi) = \bar \psi \wedge \gamma_5 \gamma \wedge D \psi \quad \rarrow \quad
    \L_{\text{RS}}(\psi) = \epsilon^{\,\mu \nu \rho \sigma} \bar \psi_\mu \gamma_5 \gamma_\nu D_\rho \psi_\sigma \,.  
\end{align}
The complete spacetime Lagrangian describing 
both the dynamics %the coupling 
of gravity (that is, of the spin-$2$ graviton field) and of the RS field $\psi$ is given by
\begin{align}
\label{sugraL}
    L_{\text{\tiny{sugra}}}(\omega^{ab},e^a,\psi) = R^{ab} \wedge e^{c} \wedge e^{d} \epsilon_{abcd} + 4 \bar \psi \wedge \gamma_5 \gamma \wedge D \psi ,
    %= R^{ab} \wedge e^{c} \wedge e^{d} \epsilon_{abcd} + 4 \bar \psi \wedge \gamma_5 \gamma_a \wedge D \psi \wedge e^a.
\end{align}
where $R^{ab}:=d\omega^{ab}+{\omega^a}_c \wedge \omega^{cb}$ is the Riemann (Lorentz) curvature 2-form
%($\omega^{ab}$ is the spin connection and $e^a$ the vielbein 1-form) 
and $\gamma:= \gamma_a e^{a}= \gamma_a {e^{a}}_{\!\mu} \, d x^{\,\mu}:=\gamma_\mu \, d x^{\,\mu}$ is the gamma-matrix 1-form.

The \emph{infinitesimal} transformation \eqref{susygaugetrRS}  is not a symmetry of the coupled theory \eqref{sugraL}, one must covariantise it %\eqref{susygaugetrRS} 
in order to maintain Lorentz covariance, that is
%\begin{equation}
\begin{align}
\label{sugrasusytr1}
    % & \psi \mapsto \psi^\upepsilon=\psi + D \upepsilon \quad \rarrow \quad \psi_\mu \mapsto \psi^\upepsilon_\mu=\psi_\mu + D_\mu \upepsilon , \\
    %& \text{infinitesimally, } 
    %\quad
    \delta_\epsilon \psi = D \epsilon \quad \rarrow \quad \delta_\epsilon \psi_\mu = D_\mu \epsilon .
\end{align}
%\end{equation}
One also needs  to  specify the following infinitesimal $\E$-transformations of $e^a$ and ${\omega^a}_b$:\footnote{See, e.g., \cite{Castellani:1991eu} and \cite{Tanii:2014gaa} for heuristic explanations.} 
% In fact, the coupled theory is quasi-invariant under the combined infinitesimal transformations\footnote{See, e.g., \cite{Castellani:1991eu} for a heuristic explanation.}
\begin{equation}
\begin{aligned}
\label{sugrasusytr2}
   % \delta_\epsilon \psi & = D \epsilon , \\
    \delta_\epsilon e^a & = i \, \b \epsilon  \, \gamma^a \psi , \\
    \delta_\epsilon \omega^{ab} & = i \, \big(  \bar D^{[a} \bar \psi^{b]} \gamma_c - 2 \, \bar D^{[a} \bar \psi_c \gamma^{b]} \big) \, \epsilon \, e^c .
\end{aligned}
\end{equation}
Under the $\E$-transformations \eqref{sugrasusytr1}-\eqref{sugrasusytr2}, the Lagrangian \eqref{sugraL} is quasi-invariant.
The field equations are
\begin{equation}
\begin{aligned}
     \label{field-eq-sugra}
    \bs E(\omega)& = 2 (De^c - \sfrac{i}{2} \bar \psi \wedge \gamma^c \psi) \wedge e^d \epsilon_{abcd} = 0, \\
     \bs E(e) &= 2 R^{ab} \wedge e^c \epsilon_{abcd} + 4 \bar \psi \wedge \gamma_5 \gamma_d D \psi = 0 , \\
      \bs E(\psi)&= 8 \gamma_5 \gamma_a D \psi \wedge e^a + 4 \gamma_5 \gamma_a \psi \wedge (De^a - \sfrac{i}{2} \bar \psi \wedge \gamma^a \psi)= 0.
\end{aligned}
\end{equation}
We remind that the supersymmetry algebra closes on-shell, that is, on the field equations.

In the supergravity literature, to get the right count of off-shell d.o.f. (namely, 12 for the gravitino), one usually ``gauge-fixes" $\psi$ by requiring the condition $\gamma^\mu \psi_\mu=0$, i.e. one restricts attention to a subspace $\Phi_o$ of the field space $\Phi$ of the theory. 
Naturally, this means one must correspondingly restricts the supersymmetry gauge group $\E$ to the subgroup $\E_o \defeq \big\{\epsilon \in \E\, |\, \slashed D \epsilon =0\, \big\}$ preserving $\Phi_o$. 
In the next section, we show that solving the functional constraint associated with this ``gauge-choice" results in a perturbative dressing.

\subsection{Gamma-trace dressing in supergravity} %%%%%%%%%%%%%%%%%%%%%%%%%%%%%%
\label{Gamma-trace dressing in supergravity}%%%%%%%%%%%%%%%%%%%%%%%%%%%%%%

We thus consider the gamma-tracelessness constraint \eqref{gammatr} as a functional condition on the variable $\psi_\mu^\upsilon \defeq \psi_\mu+ \updelta_\upsilon \psi$ and solve it explicitly for  the linear parameter $\upsilon$:
\begin{equation}
\begin{aligned}
\label{gamma-tr-dressing-sugra}
    & \gamma^{\,\mu} \psi^\upsilon_\mu = \gamma^{\,\mu} (\psi_\mu + D_\mu \upsilon) = 0 , \\
    & \Rightarrow \quad \upsilon[\psi] = - \slashed{D}\- (\gamma^{\,\mu} \psi_\mu) = - n \slashed{D}\- \chi.
\end{aligned}
\end{equation}
One checks that $\upsilon[\psi]$ satisfies \eqref{pert-dressing-field}, neglecting higher order terms. Indeed,  
given $\delta_\epsilon \, \chi = \frac{1}{n} \slashed{D} \epsilon$, we have
\begin{align}
    \delta_\epsilon \upsilon[\psi]= 
     - n \slashed{D}\- (\delta_\epsilon \, \chi ) 
     - n \cancelto{\,\text{\tiny{neglect}}}{\delta_\epsilon  (\slashed{D}\-) \chi}\!\!\!\!\!
     \approx - \epsilon.
\end{align}
One therefore builds the perturbatively dressed gravitino field
\begin{align}
    \psi^\upsilon \defeq \psi + D\upsilon[\psi] 
    = \psi - n D \slashed{D}\- \chi,
\end{align}
which is by construction gamma-traceless, $\gamma^{\,\mu} \psi_\mu^\upsilon\equiv0$, and $\E$-invariant at first order, $\delta_\epsilon \psi^\upsilon \approx 0$.% and non-local.
This means that what is commonly referred to as the gravitino field is, in fact, a (self-)dressed, \emph{non-local} field, carrying 12 (relational) d.o.f. off-shell (8 coming from its transverse spin-$\sfrac{3}{2}$ component and 4 from the spin-$\sfrac{1}{2}$ divergence field).

\subsection{Covariant transverse dressing}
\label{Covariant transverse dressing}

We now solve the functional constraint \eqref{transvcond} on the variable $\psi_\mu^\upsilon \defeq \psi_\mu+ \updelta_\upsilon \psi$  for  $\upsilon$:
\begin{equation}
\begin{aligned}
\label{div-dressing-sugra}
    & D^{\,\mu} \psi^\upsilon_\mu = D^{\,\mu} (\psi_\mu + D_\mu \upsilon) = 0 , \\
    & \Rightarrow \quad \upsilon[\psi] = - \Box\- (D^{\,\mu}\psi_\mu) .
\end{aligned}
\end{equation}
One easily checks that 
\begin{align}
    \delta_\epsilon \upsilon[\psi] =  -  \Box\- (D^{\,\mu} \delta_\epsilon(\psi_\mu)) - \cancelto{\,\text{\tiny{neglect}}}{\delta_\epsilon (\Box\- D^{\,\mu}) \psi_\mu} \!\!\!\!\!\approx -\epsilon,
\end{align}
again neglecting higher order terms.
We can then build the perturbatively dressed, \emph{non-local} gravitino field,
\begin{align}
    \psi^\upsilon := \psi + D\upsilon[\psi] = \psi - D[\Box\- (D^{\,\mu}\psi_\mu)] ,
\end{align}
which is by construction divergence-free, $D^{\,\mu} \psi_\mu^\upsilon\equiv0$, and $\E$-invariant at first order, $\delta_\epsilon \psi^\upsilon \approx 0$.
Remark that $\psi^\upsilon=\psi^\text{T}$, namely it is transverse and carries 12 d.o.f. off-shell, as it can be easily checked by considering the covariant version of \eqref{TLdivergdec}, i.e. $\psi=\psi^\text{T}+\psi^\text{L}=\psi^\text{T}+D_\mu[\Box\-(D^{\,\nu}\psi_\nu)]$.

\subsection{Dressed supergravity theory and its dynamics}
\label{Dressed supergravity theory and its dynamics}

In both the aforementioned cases, the perturbatively dressed vielbein and spin connection 1-forms read, respectively,
\begin{equation}
\begin{aligned}
\label{dressed-vielbeinandspinconn}
    (e^{a})^\upsilon &:= e^a + i \, \b \upsilon[\psi]  \, \gamma^a \psi , \\
    (\omega^{ab})^\upsilon &:= \omega^{ab} + i \, \big(  \bar D^{[a} \bar \psi^{b]} \gamma_c - 2 \, \bar D^{[a} \bar \psi_c \gamma^{b]} \big) \, \upsilon[\psi] \, e^c .
\end{aligned}
\end{equation}
According to the DFM 
\eqref{pert-dressed-Lagrangian}, with either dressing fields,
the Lagrangian $4$-form of the dressed theory is thus
\begin{align}
\label{pert-dressed-SUGRALagrangian}
  L_{\text{\tiny{sugra}}}(\omega^\upsilon, e^\upsilon, \psi^\upsilon)\defeq L_{\text{\tiny{sugra}}}(\omega, e, \phi) + d\beta(\omega, e, \psi; \upsilon). 
\end{align}
is  $\E$-invariant at first order because $\psi^\upsilon$, $\omega^\upsilon$ and $e^\upsilon$ are  supersymmetry singlet ($\E$-invariant) at first order.
Remark that these perturbatively dressed fields are \emph{relational variables} \cite{JTF-Ravera2024c}: they represents the physical, invariant relations among the (off-shell) d.o.f. of $\omega$, $e$ and $\psi$. 

The dressed field equations are
\begin{equation}
\begin{aligned}
     \label{dressed-SUGRAfield-eqs}
    \bs E(\omega^\upsilon)& = 2 [D^\upsilon(e^c)^\upsilon - \sfrac{i}{2} \bar \psi^\upsilon \wedge \gamma^c \psi^\upsilon] \wedge (e^d)^\upsilon \epsilon_{abcd} = 0, \\
     \bs E(e^\upsilon) &= 2 (R^{ab})^\upsilon \wedge (e^c)^\upsilon \epsilon_{abcd} + 4 \bar \psi^\upsilon \wedge \gamma_5 \gamma_d D^\upsilon \psi^\upsilon = 0 , \\
      \bs E(\psi^\upsilon)&= 8 \gamma_5 \gamma_a D^\upsilon \psi^\upsilon \wedge (e^a)^\upsilon + 4 \gamma_5 \gamma_a \psi^\upsilon \wedge [D^\upsilon(e^a)^\upsilon - \sfrac{i}{2} \bar \psi^\upsilon \wedge \gamma^a \psi^\upsilon]= 0.
\end{aligned}
\end{equation}
The dressed field equations \eqref{dressed-SUGRAfield-eqs} are \emph{deterministic},
they uniquely determine the evolution of the \emph{relational} d.o.f. of the theory.

\section{Conclusion}
\label{Conclusion}

We have identified instances of the DFM in elementary, yet foundational, models of supersymmetric field theories, showing that popular gauge-fixing conditions, once explicitly solved, end-up producing dressing (super)fields. 
This entails that one may build the Rarita-Schwinger spinor-vector field and the gravitino  as (non-local) supersymmetry-invariant \emph{relational} field variables, with the expected number of off-shell degrees of freedom. 
It appears that several gauge-fixings in the literature turn out to be instances of dressings (i.e. naturally construed as such). 

Remark that both the dressings here implemented in supergravity are \emph{infinitesimal}. 
One may then be interested in working out the \emph{finite} version, in a suitable space. 
The latter can be either directly spacetime or, as  appears  more natural in the context of supergravity, \emph{superspace} $M^{n|\n \mathcal{N}}$,\footnote{We denote the number of spinorial dimensions by $\n$, which, in $n=4$ spacetime dimensions, is $\n=2^{n/2}=4$.} spanned by bosonic and Grassmannian coordinates $\lbrace{x^{\,\mu},\theta^{\,\alpha}\rbrace}$. 
A particularly well-adapted approach to supergravity -- and, in general, to supersymmetric field theories --  in superspace is the so-called \emph{rheonomic approach} \cite{Neeman-Regge1978,Neeman-Regge1978b,Castellani:1991eu}, which can be seen as based on the supersymmetric extension of the Cartan geometric framework for gravity. 
For a comprehensive review of the rheonomic approach we refer the reader to \cite{DAuria:2020guc,Castellani:2019pvh,Andrianopoli:2024qwm}, and for its formal and explicit connection with Cartan supergeometry to \cite{JTF-Ravera2024review}.
In \cite{Castellani:1991eu} it was further observed that the supersymmetry transformations on spacetime can be interpreted as spanning the algebra of spacetime diffeomorphisms supplemented by super-Poincaré gauge transformations with \emph{field-dependent} parameters 
-- that is, local supersymmetry transformations are not gauge supersymmetry transformations. 
This is %even better appreciated 
clearer in superspace, where local supersymmetry transformations are seen as (infinitesimal) superdiffeomorphisms along the $\theta$-directions of superspace.
In appendix \ref{Dressing superfield in superspace} we suggest a dressing \emph{superfield} within the rheonomic approach to supergravity in superspace.

The results presented in this paper may be also relevant in the context of ``unconventional supersymmetry" \cite{Alvarez:2011gd} and its connections with supergravity \cite{Andrianopoli:2018ymh,Andrianopoli:2020zbl}. This aspect, together with the implications of the presence of the dressing fields $u$/$\upsilon$ in the boundary term of the dressed Lagrangian theory -- especially in relation with the ``boundary problem" in supergravity in the geometric approach, see \cite{Andrianopoli:2014aqa,Andrianopoli:2021rdk} -- will be analysed elsewhere.

\medskip

\section*{Acknowledgment}

J.F. is supported by the Austrian Science Fund (FWF), \mbox{[P 36542]} and by the OP J.A.C. MSCA grant, number CZ.02.01.01/00/22\_010/0003229, co-funded by the Czech government Ministry of Education, Youth \& Sports and the EU. 
L.R. acknowledges support from the COST Action CaLISTA CA21109 supported by COST.

\appendix

\section{Dressing superfield in superspace}
\label{Dressing superfield in superspace}

In this section we provide a candidate for a possible dressing superfield within the rheonomic approach \cite{Castellani:1991eu} to supergravity in superspace, considering the extension of the functional constraint \eqref{gammatr} to superfields in superspace. The latter, in fact, is the gauge-choice commonly adopted in the supergravity literature when dealing with the gravitino field to get 12 d.o.f. off-shell. The gravitino superfield 1-form reads
\begin{align}
\label{superfieldeq}
    \psi^{\,\alpha}(x,\theta) = {\psi^{\,\alpha}}_\mu (x,\theta) \,d x^{\,\mu} + {\psi^{\,\alpha}}_\beta (x,\theta)\, d \theta^{\,\beta} , 
\end{align}
where ${\psi^{\,\alpha}}_\mu (x,\theta)$ and $ {\psi^{\,\alpha}}_\beta (x,\theta)$ are the superfield components, constituting an infinite tower of fields at different order in the $\theta$-expansion. The infinitesimal transformation under (even) superdiffeomorphisms of $M^{4|4}$ of $\psi^{\,\alpha}(x,\theta)$ is written in terms of the Lie derivative $\ell_X \psi:= d(\iota_X \psi) + \iota_X (d\psi) $ along an even vector superfield $X$.
For $\mathcal{N}=1$ on-shell supersymmetry, it reduces to the simple expression
\begin{align}
\label{superspace-susytr}
    \ell_X \psi = D \epsilon =: \delta_\epsilon \psi ,
\end{align}
where $\epsilon=\epsilon^{\,\alpha}=\epsilon^{\,\alpha}(x,\theta)=\iota_X \psi^{\,\alpha}=X^{\,\mu} {\psi^{\,\alpha}}_\mu + X^{\,\beta} {\psi^{\,\alpha}}_\beta$ is the local supersymmetry (super)parameter. 

Let us now propose the following ``gamma-trace decomposition" of the super 1-form $\psi^{\,\alpha}(x,\theta)$:
\begin{equation}
\begin{aligned}
\label{superdec}
    \psi^{\,\alpha}(x,\theta) & = \big[{\uprho^{\,\alpha}}_{\,\mu}(x,\theta)+ {(\gamma_\mu)^{\,\alpha}}_\beta \, \chi^{\,\beta} (x,\theta) \big] \, dx^{\,\mu} + \big[ {\uprho^{\,\alpha}}_{\,\beta}(x,\theta)+ {(\gamma_\mu)^{\,\alpha}}_\beta \, \chi^{\,\mu} (x,\theta) \big] \, d\theta^{\,\beta} \\
    & = {\uprho^{\,\alpha}}_{\,\mu}(x,\theta) \, dx^{\,\mu} + {\uprho^{\,\alpha}}_{\,\beta}(x,\theta) \, d\theta^{\,\beta} + {(\gamma_\mu)^{\,\alpha}}_\beta \big[ \,\chi^{\,\beta} (x,\theta) \, dx^{\,\mu}  +  \chi^{\,\mu} (x,\theta) \, d\theta^{\,\beta}\,  \big] ,
\end{aligned}
\end{equation}
with ${(\gamma^{\,\mu})^\alpha}_\beta \,{\uprho^{\,\beta}}_\mu(x,\theta)=0$ and ${(\gamma_{\nu})^\alpha}_\beta \,{\uprho^{\,\beta}}_\alpha(x,\theta)=0$, and
where $\chi^{\,\beta} (x,\theta)$ is a spin-$\sfrac{1}{2}$ superfield, while $\chi^{\,\mu} (x,\theta)$ is a 4-vector (spin-$1$) superfield. Together, they form a set of 8 superfields. They are defined as
\begin{equation}
\begin{aligned}
\label{superdef}
\chi^{\,\alpha} (x,\theta) &:= \frac{1}{n} \,{(\gamma^{\,\mu})^{\,\alpha}}_\beta \, {\psi^{\,\beta}}_\mu (x,\theta) , \\
\chi^{\,\mu} (x,\theta) &:= \frac{1}{\n} \,{(\gamma^{\,\mu})^{\,\alpha}}_\beta \,{\psi^{\,\beta}}_\alpha (x,\theta) ,
\end{aligned}
\end{equation}
where $\n=2^{n/2}$ is the number of spinorial dimensions. In fact, we have
\begin{equation}
\begin{aligned}
& {(\gamma^{\,\mu})^{\,\zeta}}_\alpha \,{\psi^{\,\alpha}}_\mu (x,\theta) = {(\gamma^{\,\mu})^{\,\zeta}}_\alpha \, {(\gamma_\mu)^{\,\alpha}}_\beta \, \chi^{\,\beta} (x,\theta) = n \, \chi^{\,\zeta}(x,\theta) , \\
& {(\gamma_{\nu})^{\,\beta}}_\alpha \,{\psi^{\,\alpha}}_\beta (x,\theta) = {(\gamma_{\nu})^{\,\beta}}_\alpha \, {(\gamma_\mu)^{\,\alpha}}_\beta \, \chi^{\,\mu} (x,\theta) = \n \, \chi_\nu(x,\theta) ,
\end{aligned}
\end{equation}
where we have used the gamma-matrix formula
\begin{align}
    {(\gamma_{\nu})^{\,\beta}}_\alpha \, {(\gamma_\mu)^{\,\alpha}}_\beta = \eta_{\mu \nu} \, {\delta^{\,\alpha}}_\alpha = \eta_{\mu \nu} \cdot \n .
\end{align}
%One may then define
%\begin{align}
%    \chi^{\,Z}(x,\theta) := \lbrace{\,\chi^{\,\alpha}(x,\theta) \,,\, \chi^\mu (x,\theta) \,\rbrace}.
%\end{align}
Considering the gamma-tracelessness constraint $\gamma \cdot \psi := {(\gamma^{\,\mu})^{\, \alpha}}_\beta \, {\psi^{\,\beta}}_Z (x,\theta)=0$ (where $Z = \lbrace{\alpha,\mu \rbrace}$) on the 1-form superfield $\psi^\alpha$ in \eqref{superfieldeq} as a functional condition on the (super)variable $\psi^\upsilon:=\psi + \updelta_\upsilon \psi$ and solving it explicitly for the linear superparameter $\upsilon$, we get
\begin{equation}
\begin{aligned}
\label{superspace-gamma-tr-dressing-sugra}
    & \gamma \cdot \psi^\upsilon = \gamma \cdot (\psi + D \upsilon) = 0 , \\
    & \Rightarrow \quad \upsilon[\psi] = - \slashed{D}\- (\gamma \cdot \psi) .
    %= - \slashed{D}\- \big[ {(\gamma^{\,\mu})^{\, \alpha}}_\beta \, {\psi^{\,\beta}}_Z (x,\theta) \big].
\end{aligned} 
\end{equation}
One checks that $\upsilon[\psi]$ is actually a \emph{dressing superfield} in superspace. Indeed, it satisfies \eqref{pert-dressing-field}, neglecting higher order terms:  
\begin{align}
    \delta_\epsilon \upsilon[\psi]= 
    - \slashed{D}\- (\gamma \cdot \delta_\epsilon \psi) 
     \approx - \epsilon.
\end{align}
Finally, we build the perturbatively dressed gravitino 1-form superfield (\emph{dressed superfield})
\begin{align}
    \psi^\upsilon \defeq \psi + D\upsilon[\psi] 
    = \psi - D \slashed{D}\- (\gamma \cdot \psi) ,
\end{align}
which is by construction gamma-traceless, $\gamma\cdot \psi\equiv0$, and $\E$-invariant at first order, $\delta_\epsilon \psi^\upsilon \approx 0$.
This is a perturbative (self-)\emph{superdressing in superspace}, where the dressing superfield is given in terms of the eight superfields %defined in 
\eqref{superdef}.

Following the rheonomic approach, restricting ourselves to spacetime, s.t. 
\begin{align}
    \psi^{\,\alpha}(x,\theta)|_{\theta=d\theta=0}=\big[  {\rho^{\,\alpha}}_\mu (x) + {(\gamma_\mu)^{\,\alpha}}_\beta \, \chi^{\,\beta} (x) \big]\, dx^{\,\mu},
\end{align}
we recover the results of section \ref{Gamma-trace dressing in supergravity} for the dressed supergravity theory.

%%%%%%%%%%%%%%%%         BIBLIO             %%%%%%%%%%%%%
{
%\Huge
%\huge
%\LARGE
%\Large
%\large
%\normalsize %(default)
\small
%\footnotesize
%\scriptsize
%\tiny
 \bibliography{superdressingbib}

\begin{thebibliography}{10}

\bibitem{Rarita:1941mf}
W.~Rarita and J.~Schwinger.
\newblock {On a theory of particles with half integral spin}.
\newblock {\em Phys. Rev.}, 60:61, 1941.

\bibitem{Valenzuela:2022gbk}
M.~Valenzuela and J.~Zanelli.
\newblock {On the spin content of the classical massless Rarita-Schwinger
  system}.
\newblock {\em SciPost Phys. Proc.}, 14:047, 2023.

\bibitem{Valenzuela:2023aoa}
M.~Valenzuela and J.~Zanelli.
\newblock {Massless Rarita-Schwinger equations: Half and three halves spin
  solution}.
\newblock {\em SciPost Phys.}, 16:065, 2024.

\bibitem{GaugeInvCompFields}
C.~Fournel, J.~Fran{\c c}ois, S.~Lazzarini, and T.~Masson.
\newblock Gauge invariant composite fields out of connections, with examples.
\newblock {\em Int. J. Geom. Methods Mod. Phys.}, 11(1):1450016, 2014.

\bibitem{Francois2021}
J.~Fran{\c c}ois.
\newblock Bundle geometry of the connection space, covariant hamiltonian
  formalism, the problem of boundaries in gauge theories, and the dressing
  field method.
\newblock {\em Journal of High Energy Physics}, 2021(3):225, 2021.

\bibitem{Francois2023-a}
J.~Fran\c{c}ois~André.
\newblock The dressing field method for diffeomorphisms: a relational
  framework.
\newblock arXiv:2310.14472 [math-ph], 2023.

\bibitem{Zajac2023}
M.~Zajac.
\newblock The dressing field method in gauge theories - geometric approach.
\newblock {\em Journal of Geometric Mechanics}, 15(1):128--146, 2023.

\bibitem{Berghofer-et-al2023}
P.~Berghofer, J.~Fran{\c{c}}ois, S.~Friederich, H.~Gomes, G.~Hetzroni, A.~Maas,
  and R.~Sondenheimer.
\newblock {\em Gauge Symmetries, Symmetry Breaking, and Gauge-Invariant
  Approaches}.
\newblock Elements in the Foundations of Contemporary Physics. Cambridge
  University Press, 2023.

\bibitem{JTF-Ravera2024c}
J.~Fran\c{c}ois and L.~Ravera.
\newblock {On the Meaning of Local Symmetries: Epistemic-Ontological
  Dialectics}.
\newblock arXiv:2404.17449 [physics.hist-ph], 2024.

\bibitem{Dirac55}
P.~A.~M. Dirac.
\newblock Gauge-invariant formulation of quantum electrodynamics.
\newblock {\em Canadian Journal of Physics}, 33:650--660, 1955.

\bibitem{Dirac58}
P.~A.~M. Dirac.
\newblock {\em The principles of Quantum Mechanics}.
\newblock Oxford University Press, 4th edn edition, 1958.

\bibitem{Francois-Berghofer2024}
P.~Berghofer and J.~Fran\c{c}ois.
\newblock {Dressing vs. Fixing: On How to Extract and Interpret Gauge-Invariant
  Content}.
\newblock arXiv:2404.18582 [physics.hist-ph], 2024.

\bibitem{Castellani:1991eu}
L.~Castellani, R.~D'Auria, and P.~Frè.
\newblock {\em {Supergravity and superstrings: A Geometric perspective. Vol. 2:
  Supergravity}}.
\newblock World Scientific Pub Co Inc, 1991.

\bibitem{Singer1978}
I.~M. Singer.
\newblock Some remark on the gribov ambiguity.
\newblock {\em Commun. Math. Phys.}, 60:7--12, 1978.

\bibitem{Singer1981}
I.~M. Singer.
\newblock The geometry of the orbit space for non-abelian gauge theories.
\newblock {\em Physica Scripta}, 24(5):817--820, nov 1981.

\bibitem{Ashtekar-Lewandowski1994}
A.~Ashtekar and J.~Lewandowski.
\newblock Differential geometry on the space of connections via graphs and
  projective limits.
\newblock {\em Journal of Geometry and Physics}, 17(3):191--230, 1995.

\bibitem{Baez1994}
J.~C. Baez.
\newblock Generalized measures in gauge theory.
\newblock {\em Letters in Mathematical Physics}, 31(3):213--223, 1994.

\bibitem{Fuchs-et-al1994}
J.~Fuchs, M.~G. Schmidt, and C.~Schweigert.
\newblock On the configuration space of gauge theories.
\newblock {\em Nuclear Physics B}, 426(1):107--128, 1994.

\bibitem{Fuchs1995}
J.~Fuchs.
\newblock {The singularity structure of the Yang-Mills configuration space}.
\newblock {\em Banach Center Publications}, 39(1):287--299, 1997.

\bibitem{JTF-Ravera2024review}
J.~Fran\c{c}ois and L.~Ravera.
\newblock {Cartan geometry, supergravity, and group manifold approach}.
\newblock arXiv:2402.11376 [math-ph], 2024.

\bibitem{Gursey1987}
F.~Gursey.
\newblock Super poincar{\'e} groups and division algebras.
\newblock {\em Modern Physics Letters A}, 02(12):967--976, 1987.

\bibitem{DeAzc-Izq}
J.~A.~De Azcarraga and J.~M. Izquierdo.
\newblock {\em Lie Groups, Lie Algebras, Cohomology and some Applications in
  Physics.}
\newblock Cambridge Monographs on Mathematical Physics. Cambridge University
  Press, 1995.

\bibitem{VanNieuwenhuizen:1981ae}
P.~Van~Nieuwenhuizen.
\newblock {Supergravity}.
\newblock {\em Phys. Rept.}, 68:189--398, 1981.

\bibitem{Leibbrandt-Richardson1992}
G.~Leibbrandt and K.~A. Richardson.
\newblock Qed in a unified axial-gauge formalism with a general gauge
  parameter.
\newblock {\em Phys. Rev. D}, 46:2578--2584, Sep 1992.

\bibitem{Francois2018}
J.~Fran\c{c}ois.
\newblock {Artificial versus Substantial Gauge Symmetries: A Criterion and an
  Application to the Electroweak Model}.
\newblock {\em {Philosophy of Science}}, 86(3):472--496, 2019.

\bibitem{Tanii:2014gaa}
Y.~Tanii.
\newblock {\em {Introduction to supergravity}}, volume~1 of {\em Springer
  briefs in mathematical physics}.
\newblock Springer, Tokyo, Japan, 2014.

\bibitem{Neeman-Regge1978}
Y.~Ne'eman and T.~Regge.
\newblock {Gravity and Supergravity as Gauge Theories on a Group Manifold}.
\newblock {\em Phys. Lett. B}, 74:54--56, 1978.

\bibitem{Neeman-Regge1978b}
Y.~Ne'eman and T.~Regge.
\newblock {Gravity and Supergravity as Gauge Theories on a Group Manifold}.
\newblock {\em Riv. Nuovo Cim}, v. 1(5):1--43, 1978.

\bibitem{DAuria:2020guc}
R.~D'Auria.
\newblock {Geometric supergravity}.
\newblock arXiv:2005.13593 [hep-th], 2020.

\bibitem{Castellani:2019pvh}
L.~Castellani, A.~Ceresole, R.~D'Auria, and P.~Frè, editors.
\newblock {\em {Tullio Regge: An Eclectic Genius}: {~From Quantum Gravity to
  Computer Play}}.
\newblock World Scientific, 9 2019.

\bibitem{Andrianopoli:2024qwm}
L.~Andrianopoli and R.~D'Auria.
\newblock {Supergravity in the Geometric Approach and its Hidden Graded Lie
  Algebra}.
\newblock arXiv:2404.13987 [hep-th], 2024.

\bibitem{Alvarez:2011gd}
P.~D. Alvarez, M.~Valenzuela, and J.~Zanelli.
\newblock {Supersymmetry of a different kind}.
\newblock {\em JHEP}, 04:058, 2012.

\bibitem{Andrianopoli:2018ymh}
L.~Andrianopoli, B.~L. Cerchiai, R.~D'Auria, and M.~Trigiante.
\newblock {Unconventional supersymmetry at the boundary of AdS$_{4}$
  supergravity}.
\newblock {\em JHEP}, 04:007, 2018.

\bibitem{Andrianopoli:2020zbl}
L.~Andrianopoli, B.~L. Cerchiai, R.~Matrecano, O.~Miskovic, R.~Noris, R.~Olea,
  L.~Ravera, and M.~Trigiante.
\newblock {$ \mathcal{N} $ = 2 AdS$_{4}$ supergravity, holography and Ward
  identities}.
\newblock {\em JHEP}, 02:141, 2021.

\bibitem{Andrianopoli:2014aqa}
L.~Andrianopoli and R.~D'Auria.
\newblock {N=1 and N=2 pure supergravities on a manifold with boundary}.
\newblock {\em JHEP}, 08:012, 2014.

\bibitem{Andrianopoli:2021rdk}
L.~Andrianopoli and L.~Ravera.
\newblock {On the Geometric Approach to the Boundary Problem in Supergravity}.
\newblock {\em Universe}, 7(12):463, 2021.

\end{thebibliography}
}

\end{document}